\documentclass[aps,prb,twocolumn,superscriptaddress,longbibliography,floatfix]{revtex4-2}

\usepackage{fontspec}
\usepackage{polyglossia}
\setmainlanguage{english}
\setotherlanguage{russian}

\defaultfontfeatures{Ligatures=TeX,Renderer=HarfBuzz}

\usepackage{amsmath,amssymb,mathtools}
\usepackage{siunitx}
\usepackage{multirow}

\usepackage{graphicx}
\graphicspath{{figs/}}
\DeclareGraphicsExtensions{.pdf,.png,.jpg}

\usepackage{hyperref}
\usepackage[nameinlink,capitalise]{cleveref}

\usepackage{xcolor}
\usepackage[normalem]{ulem}

\crefname{figure}{Figure}{Figures}
\crefname{table}{Table}{Tables.}
\Crefname{figure}{Fig.}{Figs.}
\Crefname{table}{Table}{Tables}

\newcommand{\AlGaAs}{Al$_x$Ga$_{1-x}$As}
\newcommand{\tlp}{$3\lambda/2$}
\newcommand{\lch}{$\lambda/4$}
\newcommand{\Eqref}[1]{Eq.~\eqref{#1}}
\newcommand{\GaAsAlAsx}{$(\text{GaAs/AlAs})_{x_{\mathrm{Al}}}$}

\begin{document}

\title{Digital-Alloy Bragg Mirrors\\
in High-Q Microcavities for Polariton Lasing}

\author{V.\,A. Stolyarov}
\email{v.stolyarov@spbu.ru}
\affiliation{Spin Optics Laboratory, St. Petersburg State University, St. Petersburg, Russia}
\author{A.\,S. Kurdyubov}
\affiliation{Spin Optics Laboratory, St. Petersburg State University, St. Petersburg, Russia}
\author{A.\,V. Trifonov}
\affiliation{Spin Optics Laboratory, St. Petersburg State University, St. Petersburg, Russia}
\author{M.\,Yu. Petrov}
\affiliation{Spin Optics Laboratory, St. Petersburg State University, St. Petersburg, Russia}
\author{I.\,V. Ignatiev}
\affiliation{Spin Optics Laboratory, St. Petersburg State University, St. Petersburg, Russia}
\author{M.\,S. Lozhkin}
\affiliation{Nanophotonics Research Center, St. Petersburg State University, St. Petersburg, Russia}
\author{S.\,A. Eliseev}
\affiliation{Nanophotonics Research Center, St. Petersburg State University, St. Petersburg, Russia}
\author{Yu.\,P. Efimov}
\affiliation{Nanophotonics Research Center, St. Petersburg State University, St. Petersburg, Russia}
\author{V.\,A. Lovtcius}
\affiliation{Nanophotonics Research Center, St. Petersburg State University, St. Petersburg, Russia}
\author{A.\,V. Kavokin}
\affiliation{Spin Optics Laboratory, St. Petersburg State University, St. Petersburg, Russia}
\affiliation{Abrikosov Center for Theoretical Physics, Moscow Center for Advanced Studies, Moscow, Russia}
\affiliation{Russian Quantum Center, Skolkovo, Moscow, Russia}


\begin{abstract}
We present an approach to the molecular beam epitaxy of high-$Q$ planar GaAs-based microcavities in which the AlGaAs high-index layers of the distributed Bragg reflectors are replaced by short-period GaAs/AlAs superlattices (digital alloys) engineered to provide the same effective Al content. This design enables a significant reduction of interface roughness, precise control of both the \lch\ optical thickness and the effective Al content, suppression of the propagation of certain structural defects, and efficient tuning of the intrinsic absorption at the polariton emission wavelength through optimization of the superlattice parameters. Using this approach, we have grown a microcavity with a low polariton lasing threshold of $P_{\mathrm{th}} \approx \SI{570 \pm 90}{\watt\per\centi\meter\squared}$ and a high experimental quality factor of $Q_{\mathrm{expt}} \approx (5.4 \pm 0.1)\times 10^4$. This value exceeds by a factor of 2 the theoretical estimate obtained within a model in which the digital alloy is replaced by a ternary AlGaAs alloy with the same effective Al content. We demonstrate that accurate modeling of the stop-band characteristics and the $Q$ factor requires incorporation of the modified electronic density of states in the superlattice, including quantum-confinement and excitonic effects.

\end{abstract}

\maketitle

\section{Introduction}

Exciton polaritons are quasiparticles that emerge in the regime of strong coupling between light (photons) and matter (excitons) in semiconductor structures \cite{Weisbuch1992, Kavokin2003, Savona1995, Kavokin2010}. Inheriting the properties of both constituents, polaritons combine the high mobility and coherence of photons with the tunability and nonlinear response of excitons \cite{Carusotto2013}. Owing to this hybrid nature, polaritons can form a macroscopically populated coherent state described by a single wave function. Such a state is commonly referred to in the literature as a polariton Bose-Einstein condensate \cite{Imamoglu1996, Kasprzak2006, Deng2002, Balili2007, Byrnes2014, Lagoudakis2008}. The polariton condensate represents a promising platform for the development of a new generation of optoelectronic devices \cite{Sanvitto2016, Ballarini2013, Liew2011}, including quantum and analog computing architectures and simulators \cite{Kavokin2022, Liew2023, Berloff2017, Moxley2021}, optical logic elements \cite{EA2013, Sannikov2024, Zasedatelev2019}, neuromorphic networks \cite{Sedov2025, Opala2019, Gan2025}, and polariton lasers \cite{Schneider2013, Zhang2022}.

A central element of polariton-based technologies is the semiconductor microcavity \cite{Kavokin2017, Vahala2003}, which consists of two distributed Bragg reflectors (DBRs) separated by a cavity of only a few optical wavelengths and an active region embedded between them. Microcavities for polariton physics have been realized in a wide variety of material systems \cite{Bellessa2024}, including nitride-based (GaN) \cite{Christopoulos2007, Jamadi2016} and oxide-based (ZnO) \cite{Jamadi2016, Li2013, Lai2013} platforms, organic microcavities \cite{KenaCohen2010, Daskalakis2014}, halide perovskites \cite{Su2021, Peng2022}, and van der Waals transition-metal dichalcogenides (TMDCs) (WS$_2$, MoS$_2$) \cite{Chernikov2014, Lackner2021}. Nevertheless, despite this diversity, traditional planar GaAs- or InGaAs-based semiconductor microcavities with AlAs/AlGaAs DBRs remain the most widely used systems for fundamental studies and continue to serve as a technologically mature and highly tunable platform. This is primarily due to the well-developed molecular beam epitaxy (MBE) growth technology \cite{Morresi2013}, the record polariton lifetimes achievable in GaAs microcavities \cite{Steger2015, Beaumariage2024}, the flexibility of the design, and the thoroughly studied excitonic properties. Despite the 30-year history of GaAs-based microcavities, their performance continues to improve. For instance, room-temperature polariton condensation has recently been demonstrated in a GaAs-based structure \cite{Alnatah2025}.

One of the central requirements for realizing polariton Bose-Einstein condensation in the GaAs system is a sufficiently high quality factor ($Q$ factor) of the microcavity ($Q \gtrsim \num{5e3}$). The $Q$ factor is defined as
\begin{equation}
  Q = \frac{\omega_0}{\Delta\omega} = \frac{E_0}{\Delta E},
  \label{eq:Q-def}
\end{equation}
where $\omega_0$ ($E_0$) is the resonance frequency (energy) of the bare photon mode (far from the anticrossing region), and $\Delta\omega$ ($\Delta E$) is its full width at half maximum (FWHM). The $Q$ factor is determined by the reflectivity of the DBRs and by absorption and scattering losses throughout the structure.

Achieving high values of $Q$ requires a large number of DBR layer pairs ($N$). For structures designed to achieve $Q_{\mathrm{expt}} \approx 5\times10^{4}$, the total number of DBR pairs needs to reach approximately $N \approx 60$, which results in a total structure thickness of $\sim 8~\si{\micro\meter}$. This, in turn, leads to technological challenges associated with the long growth duration, as well as to the accumulation of defects that ultimately limits the achievable $Q$ factor. Besides the number of DBR periods, another fundamental parameter is the refractive-index contrast between the alternating layers, which directly determines the reflectivity of each interface and thus the attainable $Q$ factor. Furthermore, the key factors limiting the $Q$ factor include (i)~the degree of interface roughness, (ii)~deviations from the designed \lch-periodicity, (iii)~the level of optical absorption at the cavity-mode energy, and (iv)~the density of structural defects \cite{Oesterle2005, Rivera1999, Michael2007, Reitzenstein2007, Gurioli2001, Jasik2008, Gacevic2018, Tikhodeev2021}. Among these defects, screw dislocations propagating through the entire heterostructure and cross-hatch dislocations arising from elastic strain relaxation in multilayer stacks \cite{Tinkler2015, Zajac2012_1, Zajac2012_2, Andrews2002} have a particularly strong impact on the $Q$ factor. They enhance photonic disorder, increase scattering, and promote localization of the polariton lasing mode.

\begin{figure}[t]
  \centering
  \includegraphics[width=\linewidth]{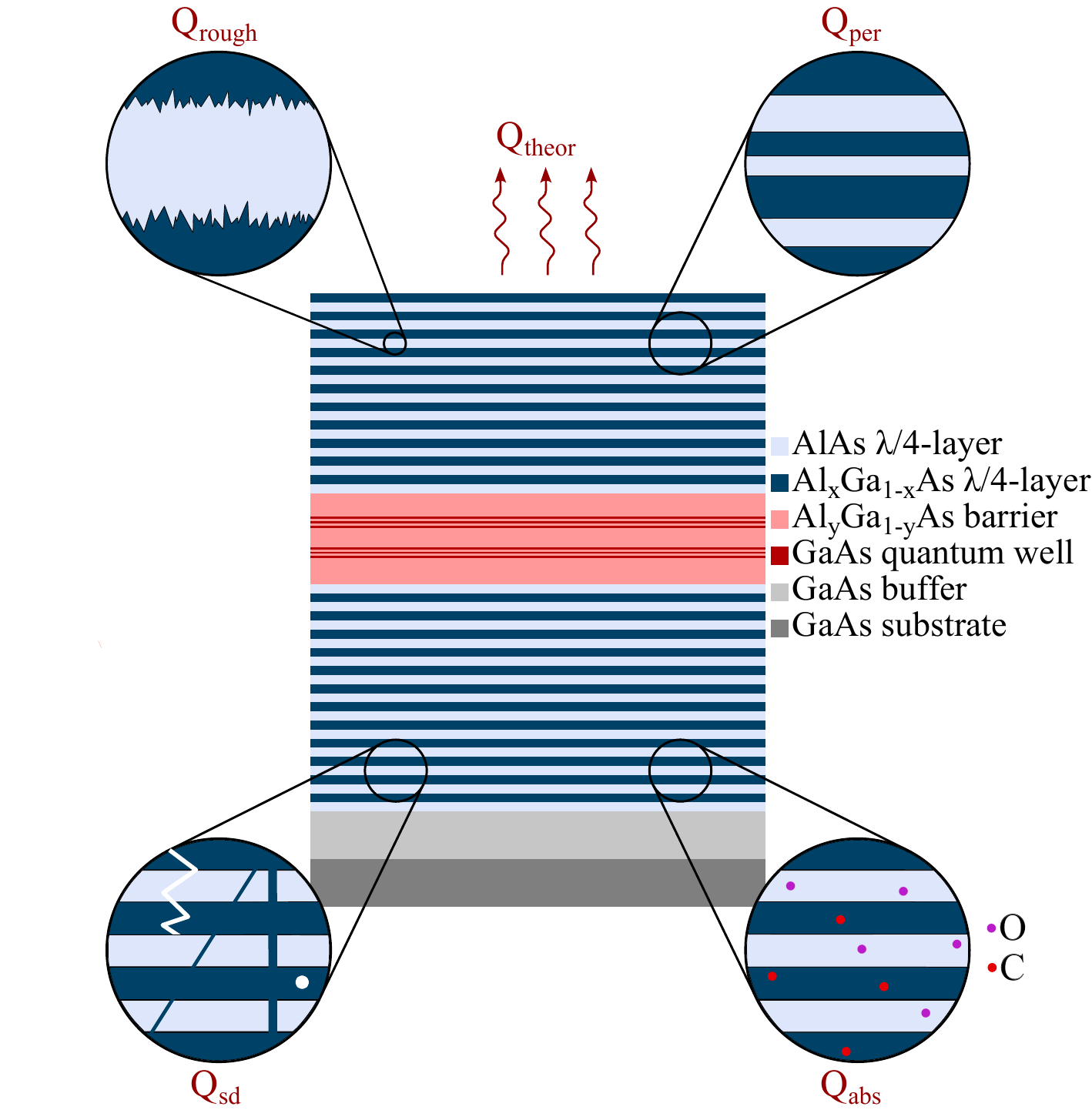} 
  \caption{Loss channels in a planar GaAs/\AlGaAs\ semiconductor microcavity that limit the achievable $Q$ factor.
$Q_{\mathrm{theor}}$ denotes radiative leakage through the mirrors.
$Q_{\mathrm{rough}}$ denotes scattering due to interface roughness.
$Q_{\mathrm{per}}$ denotes losses associated with deviations from the \lch-periodicity in the DBRs.
$Q_{\mathrm{sd}}$ denotes scattering from structural defects.
$Q_{\mathrm{abs}}$ denotes intrinsic absorption in the constituent materials.}
  \label{fig:losses}
\end{figure}

The main loss channels leading to a reduction of the $Q$ factor are schematically illustrated in \Cref{fig:losses}. For high-$Q$ microcavities, where the losses are small and effectively independent, the experimental $Q$ factor ($Q_{\mathrm{expt}}$) can be expressed as
\begin{equation}
  \frac{1}{Q_{\mathrm{expt}}} =
  \frac{1}{Q_{\mathrm{theor}}} +
  \frac{1}{Q_{\mathrm{abs}}} +
  \frac{1}{Q_{\mathrm{rough}}} +
  \frac{1}{Q_{\mathrm{per}}} +
  \frac{1}{Q_{\mathrm{sd}}}.
  \label{eq:Qsum}
\end{equation}
Here, $Q_{\mathrm{theor}}$ is the theoretical value limited solely by radiative leakage through the DBRs, $Q_{\mathrm{abs}}$ accounts for intrinsic absorption within the materials of the microcavity and cavity spacer, $Q_{\mathrm{rough}}$ corresponds to losses induced by scattering from interface roughness, $Q_{\mathrm{per}}$ describes losses caused by deviations from the intended \lch-periodicity in the DBR, and $Q_{\mathrm{sd}}$ represents scattering from structural defects.

For microcavities with GaAs quantum wells, achieving high performance typically requires the growth of high-quality ternary alloy (TA) \AlGaAs\ layers. This is a nontrivial technological problem, since the two-dimensional surface mobilities of Ga and Al atoms differ strongly at temperatures typical for MBE growth, which hinders uniform layer-by-layer epitaxy \cite{Berger1987, Asom1991}. As a consequence, surface micro-roughness, local compositional fluctuations, spatial variations of the refractive index, and an increased density of crystal defects emerge. In thick multilayer structures these inhomogeneities accumulate and lead to enhanced optical scattering, thereby limiting the microcavity $Q$ factor. Moreover, surface roughness constitutes a major source of potential fluctuations for polaritons, acting as a primary origin of photonic disorder \cite{Gurioli2001, Abbarchi2012}. 

To avoid the need for ternary \AlGaAs\ layers in microcavity DBRs, the GaAs quantum wells can be replaced by InGaAs wells. The redshift of the quantum-well exciton resonance allows the photon mode to be positioned at lower energies as well. In this spectral range, GaAs can serve as the high-index layer of the DBR, making it possible to employ fully binary AlAs/GaAs mirrors instead of AlAs/\AlGaAs\ ones. However, in both cases the structure becomes significantly strained due to InAs and GaAs lattice mismatch, resulting in a high density of cross-hatch dislocations \cite{Andrews2002, Tinkler2015}. One established method to suppress such dislocations relies on introducing phosphorus to create strain-compensating AlAsP layers or thin AlP insertions, which substantially reduces dislocation density \cite{Cilibrizzi2014, Zajac2012_3}. These approaches, however, require more complex multi-chamber MBE systems and significantly complicate the overall growth process.

In this paper, we present experimental and theoretical results for GaAs-based microcavities grown using an alternative approach. The key idea is to replace the ternary \AlGaAs\ layers with an optically equivalent short-period superlattice (SPSL) GaAs/AlAs with a period of only a few monolayers. This approach, often referred to as a digital alloy, is a standard and robust method in the MBE growth of semiconductor heterostructures \cite{Geisselbrecht1999, Hong2002, Kruse2011, Sun2017, Lyu2018, Xie2019, Kumar2023, Lei1991, Peters1993, Newman2000, Pickrell2000, Li2004, Ju2018, Cho2010}. For example, digital-alloy approaches have been extensively explored in the context of vertical-cavity surface-emitting lasers (VCSELs). In particular, digital alloys and digitally graded superlattices were used to reduce the electrical resistance of DBRs as well as associated device heating and improve carrier transport across heterointerfaces \cite{Lei1991, Peters1993, Newman2000}. In addition, such structures were shown to facilitate selective oxidation and enhance the structural stability of high-Al-content layers, which is essential for reliable oxide-aperture formation \cite{Pickrell2000, Li2004, Ju2018}. Furthermore, due to the discrete nature of binary layer deposition, digital alloys provide improved control over the effective composition, leading to enhanced wafer-scale uniformity and reproducibility compared to conventional ternary alloys \cite{Cho2010}. As a result, digital-alloy DBRs have become a well-established tool for improving manufacturability and yield in VCSEL technologies.

In contrast, for polaritonic microcavities the central challenge is not electrical, thermal, or oxidation-related, but the suppression of optical losses that limit the cavity $Q$ factor. In this regime, the digital-alloy approach provides a qualitatively different advantage. It enables a significant reduction of the technological challenges and structural imperfections described above. First, precise control of the SPSL layer thickness during growth is considerably simpler than maintaining the composition and growth rate of a ternary alloy. Second, the use of SPSLs helps suppress the propagation of threading dislocations through the structure \cite{Bedair1986, Park2020, Shinohara1985}. It should be noted, however, that its influence on the density of cross-hatch dislocations remains unclear, as indicated by Nomarski microscopy. Third, during SPSL growth, short growth interruptions can be introduced after each GaAs layer. Due to the high surface mobility of Ga atoms, these pauses smooth the heterointerface, effectively suppressing roughness before the next deposition step \cite{Berger1987, Tsuji1999}. AFM measurements of our SPSL-based samples confirmed an order-of-magnitude reduction in surface roughness compared to structures grown with ternary alloys. In addition, we show that this approach enables efficient control of absorption within the DBRs by tuning the SPSL period and its effective refractive index, which further increases the attainable $Q$ factor. Using this method, a microcavity exhibiting a quality factor of $Q_{\mathrm{expt}} \approx (5.4 \pm 0.1)\times 10^4$ was grown.

The paper is organized as follows. In Sec.~\ref{sec:methods}, the main experimental and theoretical techniques (molecular beam epitaxy, transfer-matrix modeling, and optical characterization) are summarized, and the fabricated samples are described. Sec.~\ref{sec:results} is divided into several subsections, each addressing one of the main loss channels introduced in \Eqref{eq:Qsum}. The section concludes with a description of the optimized ultra-high-$Q$ microcavity structure. We further present the $Q$ factor measurements, demonstrate polariton lasing, and discuss a phenomenological model that accounts for the effect of the superlattice within the DBR. Sec.~\ref{sec:comparison} provides a comparative analysis of the obtained microcavity characteristics with previously published results for state-of-the-art GaAs-based microcavities and alternative material systems.

\section{Methods}
\label{sec:methods}

\textbf{Samples.}  
The series of structures investigated in this work was grown by molecular beam epitaxy on GaAs(100) substrates. The growth was carried out at a constant substrate temperature of approximately \SI{580}{\celsius}. The V/III beam equivalent pressure (BEP) ratio was maintained at about 2:1. During DBR growth, the average growth rate, determined from reflection high-energy electron diffraction (RHEED) oscillations, was roughly one monolayer per second ($\sim$\,\SI{0.2827}{\nano\meter\per\second}), whereas for the cavity region the growth rate was reduced to 0.5~monolayers per second. The BEP values were also decreased by a factor of two for the cavity region while keeping the same 2:1 ratio. To obtain heterointerfaces with minimal roughness, growth interruptions of 10~seconds were introduced after each GaAs layer, while a longer 30-second interruption was applied after each quantum well. During these pauses, the Ga flux was shuttered, and in the presence of an arsenic overpressure, surface Ga atoms, owing to their high mobility, relaxed and rearranged into a more ordered configuration. During DBR growth, the substrate was rotated at approximately 20~rpm to improve the lateral uniformity of the layers across the wafer. Conversely, during growth of the cavity, the substrate rotation was stopped, resulting in a detuning gradient of approximately $\sim$\,\SI{2}{\milli\electronvolt\per\milli\meter} across the sample. A more detailed description of the technological procedures and growth parameters is provided in Appendix~A.

\begin{table}[t]
  \caption{
  Parameters of the investigated heterostructures.
  Notation: DBR (bottom/top) is the number of layer pairs in the bottom and top mirrors;
    Layer type is the type of the high-index layer in the DBR (short-period superlattice or ternary alloy);
  $d_{\mathrm{GaAs}}$ is the GaAs layer thickness in the SPSL (the AlAs layer thickness in the SPSL was fixed at $d_{\mathrm{AlAs}}=\SI{1.41}{\nano\meter}$ for all structures);
  $Q$ is the experimental $Q$ factor of the microcavity (the uncertainty of $Q$ is determined from the fitting error of the linewidth of the lower polariton branch).
  }
  \label{tab:samples}
  \begin{ruledtabular}
  \begin{tabular}{lcccc}
    Sample & DBR ($\downarrow/\uparrow$) & Layer type & $d_{\mathrm{GaAs}}$ (nm) & $Q$ \\[2pt]
    \hline
    \noalign{\vskip 2pt}
    MC1  & 23/18 & SPSL & 5.1 & $(5.6 \pm 0.3)\times 10^3$ \\
    MC2  & 35/30 & SPSL & 4.0 & $(5.4 \pm 0.1)\times 10^4$ \\
    HMC1 & 20/0  & SPSL & 5.1 & -- \\
    HMC2 & 20/0  & SPSL & 4.0 & -- \\
    HMC3 & 41/0  & TA   & --  & -- \\
  \end{tabular}
  \end{ruledtabular}
\end{table}

Several structures were fabricated for this study: two \tlp\ microcavities (MC1, MC2), each incorporating two groups of three GaAs quantum wells placed at the antinodes of the standing electromagnetic field, and three half-microcavities (HMC1, HMC2, and HMC3) consisting of a bottom DBR with a single GaAs quantum well grown on top. This design enables optical characterization of the emitting states that would otherwise be hidden beneath the stop band of the top DBR in a microcavity. The parameters of the structures are summarized in \Cref{tab:samples}. 

In all samples, the DBRs consist of alternating \lch-layers with different refractive indices. The low-index material is AlAs, with a nominal thickness of \SI{70}{\nano\meter} for MC1, HMC1, HMC2, and HMC3, and \SI{66}{\nano\meter} for MC2. The high-index layer is either a short-period GaAs/AlAs superlattice (digital alloy) (for MC1, MC2, HMC1, and HMC2) or a ternary \AlGaAs\ alloy (for HMC3).

In the case of digital-alloy DBRs, the \lch\ optical thickness was realized by stacking multiple short-period GaAs/AlAs bilayers. For MC1 and HMC1, each high-index layer consists of 8.5 periods of a superlattice with layer thicknesses of \SI{5.1}{\nano\meter} (GaAs) and \SI{1.41}{\nano\meter} (AlAs), resulting in a total thickness of approximately \SI{57.2}{\nano\meter} and an average aluminum content of $x_{\mathrm{Al}} \approx 0.2$. For MC2 and HMC2, the high-index layer is formed by 10.5 periods of a superlattice with GaAs and AlAs thicknesses of \SI{4.0}{\nano\meter} and \SI{1.41}{\nano\meter}, respectively, yielding a total thickness of approximately \SI{58.1}{\nano\meter} and an average aluminum content of $x_{\mathrm{Al}} \approx 0.24$. The physical thickness of the cavity in MC1 and MC2 corresponds to \tlp, which is approximately \SI{346}{\nano\meter}.

\textbf{Optical modeling of the structures.}  
To simulate the reflectance spectra and the electromagnetic field distribution in the microcavities, we employed a transfer-matrix method \cite{BornWolf1999}. The heterostructure was represented as a sequence of planar layers characterized by complex refractive indices $\tilde{n}_i(\lambda, T) = n_i + i k_i$ and thicknesses $d_i$, for which the corresponding transfer matrices $M_i^{\mathrm{layer}}$ and $M_i^{\mathrm{interface}}$ were constructed. In modeling the reflectance spectra, we explicitly accounted for the excitonic response of the quantum wells by introducing a dedicated matrix $M^{\mathrm{QW}}$ within the framework of the nonlocal dielectric-response formalism \cite{Ivchenko2005, Kavokin2017}. This approach replaces the standard treatment based on an effective resonant dielectric permittivity. The matrix $M^{\mathrm{QW}}$ incorporates the intrinsic parameters of the exciton resonance: the radiative ($\Gamma_0$) and nonradiative ($\Gamma$) linewidth broadenings, the exciton transition frequency ($\omega_0$), and the phase of the excitonic contribution ($\phi$). A more detailed description of the algorithm is provided in Appendix~B.

\textbf{Optical characterization and $Q$ factor measurements.}  
Optical characterization of the samples was performed using reflectance and photoluminescence (PL) spectroscopy at \SI{3.5}{\kelvin} in a continuous-flow helium cryostat. Reflectance (PL) spectra were recorded using a white-light source (a tunable cw Ti:Sa laser) and a spectrometer with a spectral resolution of approximately \SI{35}{\micro\eV}, which allowed us to resolve the stop-band width and the exciton resonances of the quantum wells. To measure the $Q$ factor, we used a tunable Ti:Sa narrow-linewidth ring-cavity laser serving as a high-precision spectroscopic probe. Continuous wavelength scanning over the 800--820~\si{\nano\meter} range with a step of about $5\times10^{-5}$~\si{\nano\meter} enabled the detection of narrow photon-mode resonances with an energy resolution of $\sim$\,\SI{1}{\micro\eV}, unattainable with standard spectrometers. To enhance measurement stability, we employed a balanced detection scheme that suppresses laser-power fluctuations. A detailed description of the experimental setup and method is provided in Appendix~C.

\section{Results and Discussion}
\label{sec:results}

\subsection{Impurity absorption ($Q_{\mathrm{abs}}$)}
\label{subsec:imp_abs}

Absorption of light by impurity-related states is one of the main loss channels limiting the theoretical $Q$ factor of microcavities. Spectrally, such transitions form a band located roughly 1--10~meV below the band gap $E_g$ of the corresponding layers, extending by several hundred meV towards lower energies. In ternary \AlGaAs\ alloys, the position of this impurity band is determined by the alloy band gap itself \cite{Aspnes1986, Gehrsitz2000, Vurgaftman2001}. Consequently, for GaAs-based quantum well microcavities, AlAs/\AlGaAs\ DBRs with $x = 0.1$--$0.3$ are typically used, ensuring that the absorption edge is sufficiently detuned from the photon-mode energy and preventing excessive damping.

\begin{figure}[t]
  \centering
  \includegraphics[width=\linewidth]{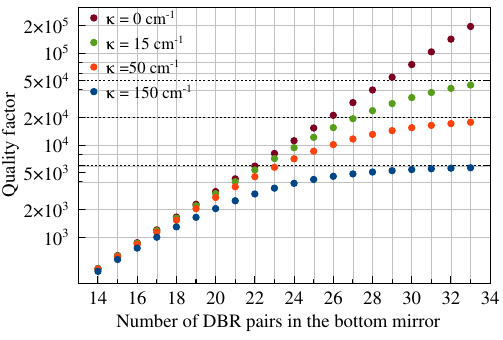}
  \caption{Calculated dependences of the microcavity $Q$ factor on the number of pairs in the bottom DBR (the top DBR contains five fewer pairs) for different values of the absorption coefficient $\kappa$.}
  \label{fig:Qfac}
\end{figure}

To evaluate the influence of impurity absorption, the dependence of the microcavity $Q$ factor on the number of $\mathrm{AlAs/Al_{0.2}Ga_{0.8}As}$ layer pairs in the bottom DBR (with the top mirror containing five fewer pairs) was calculated. The absorption was included in the model as an imaginary contribution to the refractive index of the AlGaAs layers (\Cref{fig:Qfac}).

\begin{figure}[b]
  \centering
  \includegraphics[width=\linewidth]{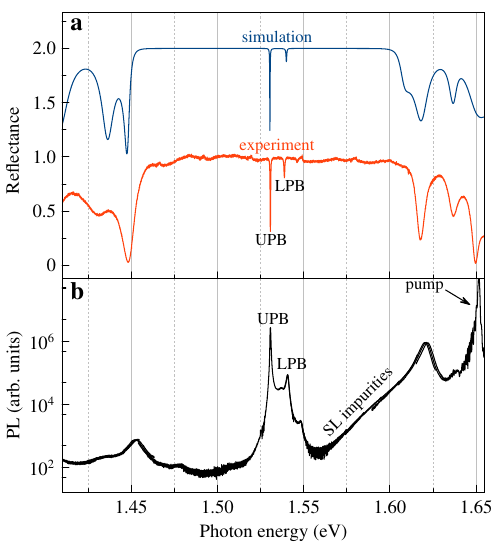}
  \caption{(a) Experimental (red) and theoretical (blue) reflectance spectra of the MC1 structure at a small negative detuning. (b) PL spectrum at small negative detuning measured under nonresonant excitation ($E_{\mathrm{pump}} = \SI{1.65}{\eV}$) showing the lower and upper polariton branches together with impurity-related transitions of the SPSL.}
  \label{fig:T888_cas}
\end{figure}

\begin{figure*}[t]
  \centering
  \includegraphics[width=\linewidth]{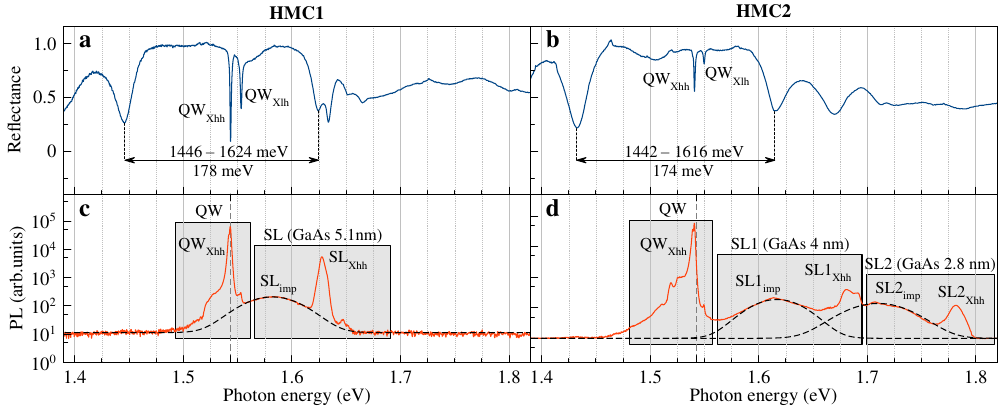}
  \caption{Optical characterization of half-microcavities with the SPSL-based DBRs and the GaAs layer thicknesses of \SI{5.1}{\nano\meter} (HMC1) and \SI{4}{\nano\meter} (HMC2). (a, b) Reflectance spectra showing the stop band of the half-microcavity and the heavy-hole ($\mathrm{QW_{Xhh}}$) and light-hole ($\mathrm{QW_{Xlh}}$) exciton transitions in the GaAs quantum well. (c, d) PL spectra under excitation at \SI{2.33}{eV}, revealing the superlattice exciton resonances ($\mathrm{SL_{Xhh}}$) and an impurity tail ($\mathrm{SL_{imp}}$).}
  \label{fig:T896_T897}
\end{figure*}

Light absorption in a material is described by the Bouguer-Lambert-Beer law \cite{BornWolf1999}
\begin{equation}
    I(x) = I_0 e^{-\kappa x},
    \label{eq:blg}
\end{equation}
where $I_0$ and $I(x)$ are the incident and transmitted intensities, respectively.  The absorption coefficient is given by \(\kappa = \frac{4\pi}{\lambda} k\) where $k$ is the imaginary part of the complex refractive index ($\tilde{n}(\lambda, T) = n + i k$), which accounts for absorption losses in the material.

As seen from the figure, the increase in $Q$ becomes strongly limited by absorption once the DBR reaches a certain thickness. Beyond this point, additional pairs contribute negligibly to the overall $Q$.

In short-period AlAs/GaAs superlattices, the position of impurity-related states is determined by the effective band-gap width of the superlattice, specifically by the energy of the excitonic miniband \cite{Cardona1987, Danan1987, Smith1990}. Size quantization of excitons (carriers) in the GaAs layers, combined with tunneling through the thin AlAs barriers, leads to the formation of minibands. Their energies, and hence the spectrum of impurity absorption, are governed by the GaAs layer thickness $d_{\mathrm{GaAs}}$ and the superlattice period (i.e., the thickness of the thin AlAs barriers). Properly choosing these layer thicknesses makes it possible to minimize the spectral overlap between impurity absorption and the microcavity photon mode, thereby significantly improving the attainable $Q$ factor.

To illustrate this effect, a microcavity sample MC1 was grown, containing 23 (18) pairs of AlAs and a short-period GaAs/AlAs superlattice layer (with an effective Al fraction of $x_{\mathrm{Al}} = 0.2$) in the bottom (top) DBR. The reflectance and PL spectra of this sample are shown in \Cref{fig:T888_cas}. The experimental $Q$ factor extracted from the reflectance spectrum using \Eqref{eq:Q-def} is $Q_{\mathrm{expt}} = (5.6 \pm 0.3)\times 10^3$. The uncertainty of $Q_{\mathrm{expt}}$ is determined from the fitting error of the linewidth of the lower polariton branch. This value corresponds to a point on the sample at large negative detuning ($\delta \approx -12~\mathrm{meV}$), where the lower polariton is predominantly photonic and the influence of exciton-photon coupling on the linewidth can be neglected. The theoretical reflectance spectrum [blue curve in \Cref{fig:T888_cas}(a)] reproduces the experimental data with high accuracy, including the stop-band position and its width as well as the Rabi splitting. The theoretical $Q$ factor was determined from the linewidth of the lower polariton branch at the same detuning and spectral position as in the experiment using \Eqref{eq:Q-def}, yielding $Q_{\mathrm{theor}} \approx \num{7e3}$. This value was obtained within a model that accounts for the fact that the refractive index of GaAs in the SPSL differs from that of bulk GaAs due to quantum-confinement and excitonic effects. This model is discussed in detail in Sec.~\ref{sec:Optimized microcavity}. Within this approach, the effect is incorporated phenomenologically by introducing an energy-dependent shift of the refractive index of the SPSL GaAs layers toward higher energies. The substantial discrepancy between experiment and theory originates from the impurity-related absorption, clearly indicating the necessity of optimizing the SPSL parameters even for relatively low-$Q$ structures with a small number of DBR periods. This reduction of the Q factor corresponds to an effective absorption coefficient of approximately $\kappa \approx 30~\mathrm{cm^{-1}}$. The conclusion regarding impurity-related absorption is supported by the PL spectrum in \Cref{fig:T888_cas}(b), measured under nonresonant excitation at $E_{\mathrm{pump}} = \SI{1.65}{\eV}$. In addition to the polariton branches, a broad PL band is observed in the range $1.5$--$1.65~\mathrm{eV}$, partially overlapping the stop band. This feature originates from impurity-related emission in the \SI{5.1}{\nano\meter}-thick GaAs layers of the SPSL in the DBRs. We note that the impurity-related spectrum in \Cref{fig:T888_cas}(b) is strongly distorted by the transmission spectrum of the top DBR.

\begin{figure*}[t]
  \centering
  \includegraphics[width=0.75\linewidth]{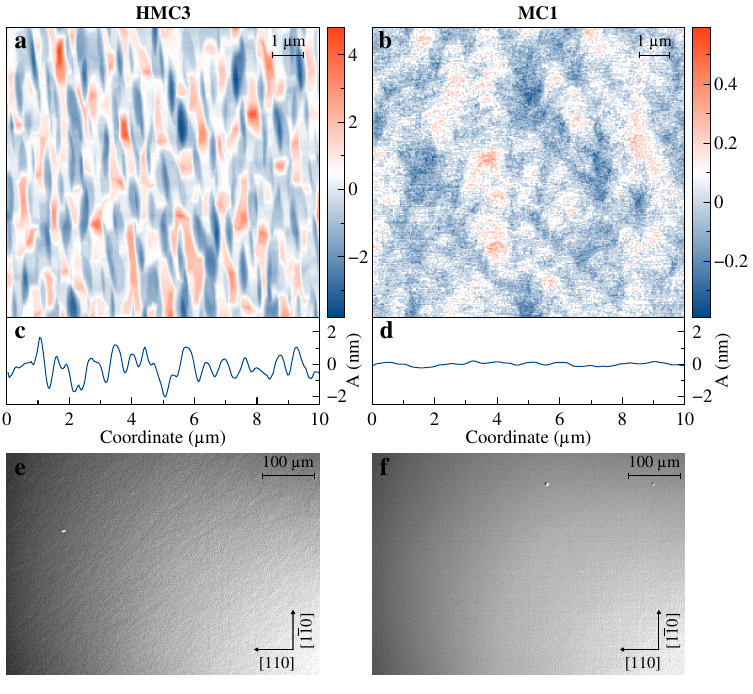}
  \caption{Atomic force microscopy (AFM) images of the surface of the HMC3 (a) and MC1 (b) samples. The HMC3 structure, which employs a ternary-alloy-based DBR, exhibits pronounced surface roughness with an RMS value of $\sim$\,\SI{1.1}{\nano\meter} (c). In contrast, the MC1 structure with an SPSL-based DBR shows a much smoother surface with an RMS value of $\sim$\,\SI{0.1}{\nano\meter} (d). Panels (e) and (f) show Nomarski micrographs of the HMC3 and MC1 samples, respectively.}
  \label{fig:AFM}
\end{figure*}

To identify optimal SPSL parameters for microcavity DBRs, two half-microcavities, HMC1 and HMC2, were fabricated, each consisting of a bottom DBR topped with a single \SI{12.5}{\nano\meter} GaAs quantum well. Such structures serve as effective test samples prior to full microcavity growth. They allow independent characterization of both the DBR (stop-band position and width) and the quantum well (PL intensity and exciton linewidth). The key difference between HMC1 and HMC2 is the GaAs-layer thickness in the SPSL ($d_{\mathrm{GaAs}}$). With respect to this parameter, HMC1 is equivalent to MC1 ($d_{\mathrm{AlAs}} = \SI{1.41}{\nano\meter}$, $d_{\mathrm{GaAs}} = \SI{5.1}{\nano\meter}$, average $x_{\mathrm{Al}} = 0.2$), whereas HMC2 corresponds to MC2 ($d_{\mathrm{AlAs}} = \SI{1.41}{\nano\meter}$, $d_{\mathrm{GaAs}} = \SI{4.0}{\nano\meter}$, average $x_{\mathrm{Al}} = 0.24$), discussed in the next subsection. To preserve the overall \lch{} optical thickness, the number of SPSL periods was increased when the GaAs-layer thickness was reduced. In the reflectance spectra of HMC1 [\Cref{fig:T896_T897}(a)] and HMC2 [\Cref{fig:T896_T897}(b)], one clearly observes the heavy-hole and light-hole exciton resonances of the GaAs quantum well, as well as the DBR stop-band edges.

A more informative set of results for our analysis is provided by the PL spectra of the HMC1 and HMC2 structures [\Cref{fig:T896_T897}(c, d)]. In both cases, we observe PL from the light-hole and heavy-hole excitons near $\sim$\,\SI{1.55}{\eV}, along with several additional spectral lines at higher energies. For HMC1, which contains an SPSL with a GaAs layer thickness of \SI{5.1}{\nano\meter}, a superlattice exciton resonance appears at $\sim$\,\SI{1.63}{\eV}, accompanied by a long low-energy impurity tail extending toward the stop band of the corresponding microcavity and even overlapping the exciton (or polariton in the full MC) resonance. As discussed earlier for MC1, such overlap limits the experimental $Q$ factor because the photon mode is efficiently absorbed in the SPSL.

In the HMC2 structure, the GaAs layer thickness of the SPSL in the DBR (SL1) is reduced to \SI{4}{\nano\meter}, shifting the superlattice exciton resonance to $\sim$\,\SI{1.68}{\eV}. This shift ensures that the broad low-energy impurity tail does not interfere with the propagation of the electromagnetic field near the exciton resonance and, in the case of the full microcavity, near the anticrossing region. Moreover, the PL spectrum of HMC2 shows an additional resonance originating from the superlattice forming the barrier of the quantum well (SL2), with GaAs and AlAs layer thicknesses of \SI{2.8}{\nano\meter} and \SI{1.41}{\nano\meter}, respectively ($x_{\mathrm{Al}} = 0.33$). This resonance appears at a higher energy, around $\sim$\,\SI{1.78}{\eV}, and also exhibits a characteristic broad low-energy impurity tail.

To quantify the overlap of the impurity-related SPSL emission with the QW spectral region, the impurity bands were fitted with Gaussian profiles [\Cref{fig:T896_T897}(c, d)], and the fraction of their spectral weight lying below a reference energy of $E_{\mathrm{ref}}~=~\SI{1.543}{\eV}$ was determined, corresponding to the typical QW exciton energy in the studied microcavity structures. Using this procedure, we obtain approximately $4.7\%$ for the SPSL with $d_{\mathrm{GaAs}} = \SI{5.1}{\nano\meter}$, $0.062\%$ for the SPSL with $d_{\mathrm{GaAs}} = \SI{4}{\nano\meter}$, and $0.00000003\%$  for the SPSL with $d_{\mathrm{GaAs}} = \SI{2.8}{\nano\meter}$. This demonstrates a strong suppression of the low-energy impurity-related spectral weight in the QW spectral region with decreasing GaAs layer thickness in the SPSL, confirming that impurity absorption can be effectively minimized by appropriate design of the digital-alloy DBRs. Optimized parameters were used in the growth of the MC2 structure discussed below.

\subsection{Interface roughness ($Q_{\mathrm{rough}}$)}

Another important factor limiting the microcavity $Q$ factor and polariton mobility is photonic disorder arising from microscopic roughness at the DBR heterointerfaces. Such irregularities lead to scattering of the photon modes and fluctuations of their local resonance energy. Replacing ternary AlGaAs layers with short-period GaAs/AlAs superlattices in the DBRs significantly suppresses the accumulation of interface roughness during the growth process \cite{Berger1987, Tsuji1999}. This improvement results from the introduction of 10-second growth interruptions after each GaAs layer, during which the surface atoms relax and the interface becomes smoother.

To compare the effect of GaAs/AlAs SPSLs with that of ternary AlGaAs on DBR interface quality, we fabricated the HMC3 structure, consisting of 41 DBR periods of alternating AlAs/Al$_{0.2}$Ga$_{0.8}$As layers. This structure serves as a direct analogue of MC1 in terms of DBR pair count and nominal alloy composition. However, in MC1 the DBR was implemented as an AlAs/\GaAsAlAsx$_{=0.2}$ SPSL.

AFM measurements revealed that the HMC3 sample exhibits pronounced surface roughness, with peak-to-valley height variations reaching $\sim$\,\SI{8.6}{\nano\meter} and an RMS roughness of \SI{1.1}{\nano\meter} [\Cref{fig:AFM}(a, c)]. In contrast, the surface of MC1 shows height variations below $\sim$\,\SI{1}{\nano\meter}, corresponding to roughly three GaAs monolayers, and an RMS roughness of only $\sim$\,\SI{0.1}{\nano\meter} [\Cref{fig:AFM}(b, d)]. Such values correspond to an atomically smooth surface. Thus, implementing short-period superlattices in the DBR structure effectively suppresses the accumulation of microscopic roughness and leads to significantly smoother heterointerfaces in the microcavity.

\begin{figure}[t]
  \centering
  \includegraphics[width=\linewidth]{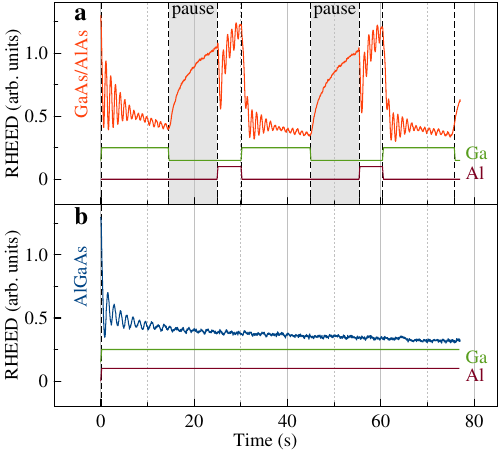}
  \caption{RHEED oscillations during the growth of a short-period GaAs/AlAs superlattice (a) and a ternary AlGaAs alloy (b). The measurements were performed during the growth of the same sample under identical growth conditions. The green (brown) curve corresponds to the opening and closing of the Ga (Al) shutters, respectively.}
  \label{fig:RHEED}
\end{figure}

\subsection{Maintaining the \lch-periodicity ($Q_{\mathrm{per}}$)} 
Another important factor determining the microcavity $Q$ factor is the precise control of the optical \lch-periodicity of the DBR layers. Even small deviations of the layer thicknesses from their nominal values can shift the spectral position of the stop band and reduce the mirror reflectivity \cite{Oesterle2005}. Therefore, the growth of multilayer structures by molecular beam epitaxy requires highly stable fluxes and precise control of the growth rates. The most reliable in situ method for monitoring the layer thickness during growth is the analysis of RHEED oscillations, which allows direct measurement of the growth rate with high accuracy. However, when depositing thick layers, these oscillations rapidly damp due to the gradual accumulation of surface roughness. This makes continuous monitoring impossible and leads to cumulative errors in the \lch-layer thickness when growing ternary-alloy layers.

\begin{figure*}[t]
  \centering
  \includegraphics[width=\linewidth]{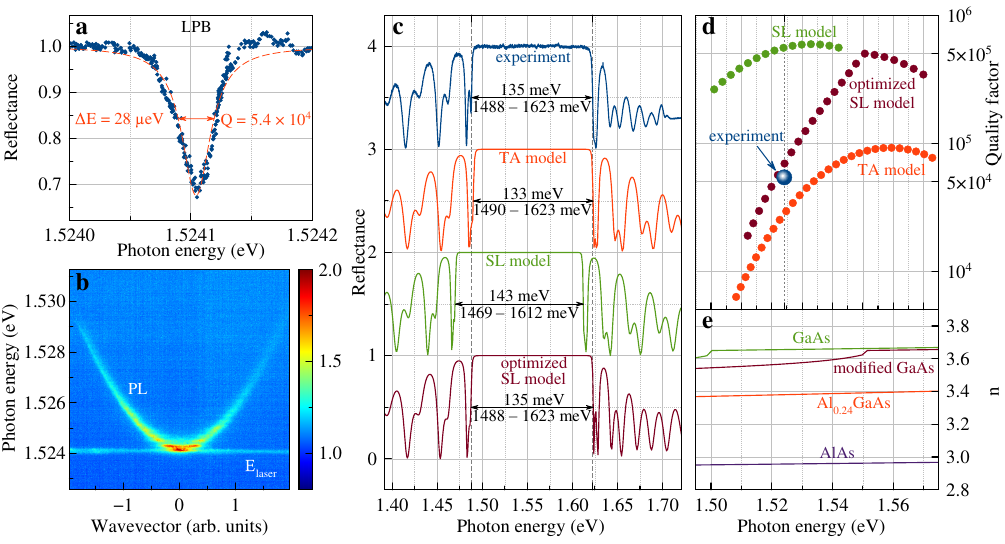}
  \caption{(a) High-resolution reflectance scan of the photon mode in sample MC2: the full width at half maximum $\Delta E \approx \SI{28.2 \pm 0.5}{\micro\eV}$ corresponds to the $Q$ factor of $Q_{\mathrm{expt}} \approx (5.4 \pm 0.1)\times 10^4$. (b) Angle-resolved PL of the MC2 microcavity under excitation at \SI{660}{\nano\meter} ($E_{\mathrm{pump}} = \SI{1.88}{\eV}$). The minimum of the lower polariton branch coincides with the energy of the scattered light from the laser tuned to the center of the reflectance dip measured by balanced photodetection. (c) Experimental reflectance spectrum (blue circle) compared with three transfer-matrix calculations: the effective-alloy TA model (red), the simple SL model (green), and the optimized SL model incorporating the modified refractive index of GaAs in the SPSL (brown). (d) Calculated $Q$ factor as a function of photon-mode energy for the three models. The experimental value (blue) lies between the TA- and SL-model predictions, while the optimized SL model provides the closest agreement. (e) Spectral dependence of the refractive index for bulk GaAs, AlAs, Al$_{0.24}$GaAs, and the modified GaAs layer in the SPSL used in the optimized model.}
  \label{fig:T906_Qfac_v2}
\end{figure*}

In contrast, for short-period GaAs/AlAs superlattices, RHEED oscillations can be maintained throughout the entire growth because the heterointerfaces remain atomically smooth, owing to the short growth interruptions introduced after each GaAs layer \cite{Berger1987}. \cref{fig:RHEED} compares RHEED oscillations during the growth of a short-period superlattice (a) and a ternary alloy (b). For the ternary alloy, the oscillations become indistinguishable after approximately 20~seconds from the onset of growth. In contrast, for the short-period superlattice, the oscillations remain clearly observable for each subsequent period due to growth interruptions introduced after the deposition of each GaAs layer. This ensures high reproducibility of the layer thicknesses and accurate control of the \lch-periodicity across the DBR stack. Moreover, the effective Al composition in the SPSL can be specified with high precision simply by setting the ratio of the GaAs and AlAs sublayer thicknesses, whereas ternary \AlGaAs\ alloys inevitably exhibit local compositional fluctuations and spatial variations of the refractive index.

\subsection{Structural defects ($Q_{\mathrm{sd}}$)}
Structural defects arising during the microcavity growth strongly influence both the $Q$ factor and the polariton transport, as they increase photonic disorder and enhance light scattering. Among these defects, threading screw dislocations and cross-hatch dislocations, formed through strain relaxation in thick multilayer structures, provide a pronounced contribution to scattering losses \cite{Tinkler2015, Zajac2012_1, Zajac2012_2, Cilibrizzi2014, Zajac2012_3}. The use of short-period GaAs/AlAs superlattices in DBRs is an established approach for suppressing the propagation of threading dislocations. The alternating thin binary layers form a dense set of heterointerfaces, each partially relaxing local strain and hindering the vertical propagation of such defects. As a result, SPSL-based DBRs generally exhibit a lower density of threading defects than those grown using ternary AlGaAs alloys \cite{Bedair1986, Park2020, Shinohara1985}.

The situation is more complex for cross-hatch dislocations, which form along the $[110]$ and $[1\bar{1}0]$ crystallographic directions during strain relaxation. Nomarski microscopy reveals characteristic cross-hatch patterns on the surfaces of both MC1 [\Cref{fig:AFM}(f)] and HMC3 [\Cref{fig:AFM}(e)]. However, a direct comparison is hindered by the pronounced surface roughness of HMC3, where height variations of several nanometers partially mask the dislocation contrast. Thus, the extent to which short-period superlattices influence the suppression of cross-hatch dislocations remains an open question and requires further dedicated investigation.

\subsection{Optimized microcavity}
\label{sec:Optimized microcavity}

\textbf{Sample.}  
As a result of the above analysis, the MC2 microcavity was designed and fabricated, incorporating 35 DBR pairs in the bottom mirror, 30 pairs in the top mirror, and two sets of three 14-nm-thick GaAs quantum wells placed at the antinodes of the \tlp-cavity. The high-index layers of the DBRs were implemented using a short-period superlattice with the following parameters: $d_{\mathrm{AlAs}} = \SI{1.41}{\nano\meter}$, $d_{\mathrm{GaAs}} = \SI{4.0}{\nano\meter}$, and an average aluminum content of $x_{\mathrm{Al}} = 0.24$. \cref{fig:T906_Qfac_v2} presents the experimental and theoretical characterization of the MC2 microcavity. The normalized reflectance spectrum is shown in \Cref{fig:T906_Qfac_v2}(c) (blue curve, shifted upward by 3 units). As seen from the figure, the large number of DBR pairs results in sharp stop-band edges. The stop-band position and width are in reasonable agreement with the transfer-matrix calculation (red curve in \Cref{fig:T906_Qfac_v2}(c), shifted upward by 2 units), where the SPSL was modeled as an optically equivalent ternary alloy with an effective concentration of $x_{\mathrm{Al}} = 0.24$.

\textbf{$\textbf{Q}$ factor.}  
Polariton resonances are not visible in \Cref{fig:T906_Qfac_v2}(c) because their linewidth is narrower than both the spectral resolution of the spectrometer and the step size used in the simulations. For this reason, the experimental $Q$ factor was determined independently using balanced photodetection with a tunable Ti:Sa narrow-linewidth ring laser, providing an effective spectral resolution below \SI{1}{\micro\electronvolt}. The results are shown in \Cref{fig:T906_Qfac_v2}(a), where the spectral feature corresponding to the lower polariton mode exhibits a $\mathrm{FWHM} \approx \SI{28.2 \pm 0.5}{\micro\eV}$, yielding $Q_{\mathrm{expt}} \approx (5.4 \pm 0.1)\times 10^4$. The uncertainties are determined from the fitting error of the linewidth of the lower polariton branch. The measurement was performed at a probe power density of approximately \SI{40}{\watt\per\centi\meter\squared} and a detuning of $\sim -12~\mathrm{meV}$. It should be noted that the measurement was performed at the most negative detuning accessible on the sample, as regions with larger negative detuning, where the measured value would represent the bare-photon mode, became effectively matte during MBE growth, resulting in strongly reduced reflectivity. Furthermore, the anticrossing region is shifted downward in energy by roughly \SI{20}{\milli\eV} relative to the center of the stop band, which also contributes to the reduction of the experimentally observed $Q_{\mathrm{expt}}$. Nevertheless, the obtained value is sufficiently high to be considered a reliable lower bound for the true $Q$ factor of MC2. 

The angle-resolved PL spectrum under nonresonant pumping at $E_{\mathrm{pump}} = \SI{1.88}{\eV}$ [\Cref{fig:T906_Qfac_v2}(b)] displays the characteristic dispersion of the lower polariton branch. Its minimum coincides with the energy of the scattered light from the narrow-linewidth laser used for obtaining results presented in \Cref{fig:T906_Qfac_v2}(a). This fact confirms that the sharp resonance observed in balanced detection corresponds to the lower polariton mode. The theoretical $Q$ factor obtained from transfer-matrix modeling is $Q_{\mathrm{theor_{TA}}} \approx \num{2.6e4}$, when the SPSL is approximated as a ternary alloy. Thus, for the MC2 microcavity, the measured $Q_{\mathrm{expt}}$ exceeds this theoretical estimate by a factor of two. The origin of this discrepancy is discussed below. 

A meaningful comparison can be made between the experimentally obtained $Q$ factor and the theoretical predictions. It should be noted that theoretical predictions discussed below take into account the deviation of the photon-mode energy from the center of the stop band, where the highest $Q$ is achieved, and neglect the interaction with the exciton. The standard approach for modeling DBRs incorporating GaAs/AlAs SPSLs is to replace the SPSL with an effective Al$_x$Ga$_{1-x}$As layer whose aluminum concentration $x$ equals the average Al content of the SPSL. The resulting theoretical spectral dependence of the $Q$ factor, calculated within this model using the MC2 structural parameters, is shown by the red points in~\Cref{fig:T906_Qfac_v2}(d). As seen from the figure, the maximum theoretical $Q$ occurs at the center of the stop band and reaches $Q \approx \num{9e4}$. For the spectral position corresponding to the experiment ($E \approx  \SI{1.5241}{\eV}$), the predicted value is $Q_{\mathrm{min}} \approx \num{2.6e4}$, which is below the experimentally measured value.

If the SPSL is modeled explicitly as a sequence of many thin GaAs and AlAs layers with the refractive indices of the corresponding bulk materials [green points in~\Cref{fig:T906_Qfac_v2}(d)], the calculated $Q$ factor increases significantly, becoming nearly an order of magnitude larger. At the experimental photon-mode energy, the predicted value is $Q_{\mathrm{max}} \approx \num{5.2e5}$. However, it should be noted that within this approach the stop band shifts to lower energies and becomes broader compared to the experiment (green curve in~\Cref{fig:T906_Qfac_v2}(c), shifted upward by 1 unit). This artificially increases the estimated $Q_{\mathrm{max}}$, as the photon mode is placed closer to the center of the incorrectly predicted stop band. The experimental $Q$ factor, which represents a lower experimental bound, lies between the two theoretical curves. Neither of these calculations agrees with the experiment because both use incorrect refractive-index values for the SPSL layers.

A more accurate modeling of the reflectance spectrum must incorporate the fact that the refractive index of the GaAs layers in the SPSL differs from that of bulk GaAs. This difference arises from the altered electronic density of states in the superlattice, where quantum-confinement and excitonic effects must be taken into account. The precise spectral dependence of the refractive index in the SPSL is a subject of separate investigation. Nevertheless, our simulations indicate that the experimentally observed stop-band width and $Q$ factor can be reproduced within a simplified phenomenological model. In this approach, the energy dependence of refractive index of GaAs in the SPSL is effectively shifted to higher energies due to the exciton quantum-confinement effect. \cref{fig:T906_Qfac_v2}(e) shows the spectral behavior of the refractive indices for bulk GaAs, AlAs, the effective AlGaAs alloy, and the SPSL-modified GaAs refractive index. As can be seen, in the spectral range 1.5–1.55~eV the dielectric contrast between AlAs and the SPSL GaAs layers becomes lower than that of bulk AlAs/GaAs. As a consequence, the $Q$ factor in this spectral region is reduced compared to the initial SPSL estimate [brown points in~\Cref{fig:T906_Qfac_v2}(d)]. The theoretical value in this model, $Q_{\mathrm{theor}} \approx \num{6.9e4}$, is only slightly higher than the experimental estimate. At the same time, the calculated reflectance spectrum reproduces the experimental one with high accuracy [brown curve in~\Cref{fig:T906_Qfac_v2}(c)]. We note that above 1.55~eV the refractive index of the SPSL GaAs in our model becomes nearly identical to that of bulk GaAs. This explains the kink in the theoretical $Q$ factor dependence at 1.55~eV in~\Cref{fig:T906_Qfac_v2}(d). The model indicates that, in the absence of excitonic interaction, the bare microcavity with parameters corresponding to MC2 can exhibit a photon-mode quality factor of up to $Q \approx 5 \times 10^{5}$ at the stop-band center. This phenomenological model was also used to calculate the Q factor and absorption coefficient $\kappa$ estimates in the MC1 sample discussed above.

\textbf{Analysis of loss mechanisms limiting the Q factor.}

Within the framework of this work, it is important to assess the relative impact of the different loss mechanisms introduced in \Eqref{eq:Qsum} on the Q factor of the microcavity. We first note that separating individual loss channels ($Q_{\mathrm{abs}}$, $Q_{\mathrm{rough}}$, $Q_{\mathrm{per}}$, $Q_{\mathrm{sd}}$) both experimentally and theoretically is a highly nontrivial task. In practice, the reduction of the experimental Q factor results from the combined action of multiple mechanisms, and unambiguous attribution to a specific contribution would require a dedicated set of samples in which only one parameter is varied at a time (e.g., absorption, periodicity, or interface roughness).

Nevertheless, a semi-quantitative analysis can be performed, which allows us to estimate the individual contributions. We emphasize that the accuracy of this analysis is limited to an order-of-magnitude estimate.

To analyze the different contributions, it is convenient to rewrite \Eqref{eq:Qsum} in terms of linewidths, as a sum of the photon-mode broadenings caused by different mechanisms:
\begin{equation}
\hslash \Gamma_{\mathrm{exp}} = \hslash \Gamma_{\mathrm{theor}} + \hslash \Gamma_{\mathrm{per}} + \hslash \Gamma_{\mathrm{rough}} + \hslash \Gamma_{\mathrm{sd}} + \hslash \Gamma_{\mathrm{abs}}.
\end{equation}

Comparing the contributions in terms of linewidths is more transparent, since the different mechanisms enter linearly and can be directly compared. This representation also allows one to estimate the theoretical limits. Below, we discuss the individual mechanisms separately.

\textit{(i) Absorption.}

We first estimate the absorption-related contribution. For the low-Q MC1 structure, we assume that the dominant reduction of the Q factor is due to absorption. Using the transfer-matrix model with an effective refractive index of the superlattice, we take $Q_{\mathrm{theor}} \approx 7 \times 10^3$, while the experimentally measured value is $Q_{\mathrm{expt}} \approx (5.6 \pm 0.3) \times 10^3$. This corresponds to an effective absorption coefficient of approximately $30~\mathrm{cm^{-1}}$. In \Cref{fig:Qfac}, the corresponding dependence would lie between the red and green data points. This, in turn, corresponds to an absorption-induced broadening of approximately $\hslash \Gamma_{\mathrm{abs}} \approx 52~\mu\mathrm{eV}$.

We further assume that the absorption scales with the low-energy spectral weight of the impurity-related SPSL emission. Using the experimentally observed suppression factor of $\sim 75$ for MC2 compared to MC1, we estimate an absorption-related broadening of $\hslash \Gamma_{\mathrm{abs}} \approx 1~\mu\mathrm{eV}$ in MC2.

Within this picture, absorption is the dominant loss mechanism in MC1, while in MC2 it is strongly suppressed.

\textit{(ii) Thickness fluctuations and (iii) interface roughness.}

Thickness fluctuations and interface roughness of the constituent layers are closely related manifestations of structural disorder and, after spatial averaging, can be treated within a common framework as fluctuations of the local optical thickness. However, fluctuations in the DBR mirrors and in the cavity spacer affect the Q factor in qualitatively different ways.

Fluctuations in the DBR mirrors lead to a reduction of reflectivity due to phase decoherence of the reflected light. In the present structures, this contribution is negligible, resulting in a linewidth broadening on the order of $\sim 0.2~\mu\mathrm{eV}$.

In contrast, fluctuations of the effective cavity spacer play a dominant role. The impact of cavity-thickness disorder depends critically on its lateral correlation length $\xi$~\cite{Duparre2002}. For very short-range disorder ($\xi \ll \lambda$), fluctuations are averaged out and the resulting losses are weak, whereas for disorder smooth on a scale much larger than the wavelength ($\xi \gg \lambda$), one expects predominantly inhomogeneous broadening.

AFM measurements (\Cref{fig:AFM}) show that the lateral scale of the roughness in our samples does not fall into either of these asymptotic limits. The measured autocorrelation function is used to estimate the lateral correlation length, yielding $\xi \approx 300$~nm for the MC1 microcavity, which is comparable to the optical wavelength in the material. This places the system in an intermediate regime.

In this regime, the resonance energy of the photon mode is determined by the average cavity thickness, while thickness fluctuations predominantly lead to scattering losses from the ideal cavity mode. In other words, the disorder does not simply induce local energy shifts, but also couples the fundamental cavity mode to finite-$k$ photonic states, thereby reducing the Q factor.

As an order-of-magnitude upper-bound estimate for disorder-induced scattering losses, one may write
\begin{equation}
\hslash \Gamma_{\mathrm{scat}} \sim E_c \left( \frac{2\pi n \sigma}{\lambda} \right)^2,
\end{equation}
in agreement with the classical result of Bennett and Porteus~\cite{Bennett1961}. Using this expression, we obtain $\hslash \Gamma_{\mathrm{scat}} \approx 200~\mu\mathrm{eV}$ for $\sigma = 1$~nm and $\hslash \Gamma_{\mathrm{scat}} \approx 2~\mu\mathrm{eV}$ for $\sigma = 0.1$~nm. A more detailed analysis of the impact of $Q_{\mathrm{rough}}$ and $Q_{\mathrm{per}}$ on the Q factor is provided in Appendix~D.

\textit{(iv) Structural defects.}

We neglect the contribution from local defects, $\hslash \Gamma_{\mathrm{sd}}$, since sufficiently large defect-free regions can be identified on the sample [see the cross-hatched defects in \Cref{fig:AFM}(f)].

Thus, for the optimized MC2 structure, we obtain the following upper-bound estimates: $\hslash \Gamma_{\mathrm{per}} + \hslash \Gamma_{\mathrm{rough}} \approx 2~\mu\mathrm{eV}$, $\hslash \Gamma_{\mathrm{sd}} \approx 0$, and $\hslash \Gamma_{\mathrm{abs}} \approx 1~\mu\mathrm{eV}$. 

For comparison, we have calculated the theoretical Q factor for the MC2 structure using the transfer-matrix method within the effective refractive index model for the superlattice, yielding $Q_{\mathrm{theor}} \approx 6.9 \times 10^4$, which corresponds to a linewidth of approximately $\hslash \Gamma_{\mathrm{theor}} \approx 22~\mu\mathrm{eV}$. Including the additional contributions estimated above ($\hslash \Gamma_{\mathrm{per}} + \hslash \Gamma_{\mathrm{rough}} + \hslash \Gamma_{\mathrm{abs}} \approx 3~\mu\mathrm{eV}$), one expects a total linewidth of $\hslash \Gamma_{\mathrm{exp}} \approx 25~\mu\mathrm{eV}$, corresponding to $Q \approx 6.1 \times 10^4$. This value is in good agreement with the experimentally measured linewidth of $\sim (28.2 \pm 0.5)~\mu\mathrm{eV}$ ($Q \approx (5.4 \pm 0.1) \times 10^4$).

Finally, we note that further increasing the number of DBR pairs leads to saturation of the Q factor. Within the present loss budget, the Q factor is expected to approach a limiting value of $Q \sim 5 \times 10^5$ for structures with a larger number of DBR pairs, evaluated at the same detuning and spectral position of the photon mode as considered above.

\textbf{Pathways toward higher Q factors.}
To achieve further increase of the Q factor, it is necessary to reduce the contributions of the main loss channels, namely $Q_{\mathrm{abs}}$, $Q_{\mathrm{per}}$, and $Q_{\mathrm{rough}}$. 

A further reduction of absorption could, in principle, be achieved by decreasing the GaAs layer thickness in the SPSL, thereby shifting the impurity-related spectral weight away from the cavity mode. However, this approach is limited by the accompanying reduction of the refractive-index contrast between the effective high- and low-index layers, which degrades the DBR reflectivity. Simultaneous reduction of both GaAs and AlAs layer thicknesses is further constrained by the minimum stable thickness of AlAs layers ($\sim 1.4$~nm), below which interface intermixing, island formation, and increased roughness may occur. In addition, further suppression of absorption requires the use of ultra-high-purity Ga sources (99.9999999\% or higher) and improved background conditions in the growth chamber to minimize impurity-related absorption.

The contribution of $\hslash \Gamma_{\mathrm{per}} + \hslash \Gamma_{\mathrm{rough}}$ is fundamentally limited by the lateral correlation properties of the disorder. In the present structures, the RMS roughness of $\sim 0.1$~nm is already below the lattice constant and therefore cannot be significantly reduced. A modest improvement in RMS roughness may be achievable by increasing the growth interruption time at GaAs interfaces, which enhances adatom surface diffusion and promotes interface smoothing. However, such an approach is expected to yield only limited improvement, while significantly increasing the total growth time and potentially introducing additional risks, such as background contamination. Therefore, the primary route toward further reduction of these losses is through control of the lateral correlation length of the disorder. To the best of our knowledge, there are no systematic studies addressing the independent control of the correlation length while maintaining a low fluctuation amplitude. General considerations suggest that growth parameters such as substrate temperature and growth rate may influence the correlation length, but this remains a subject for future technological investigation.

The contribution of structural defects is negligible in the present measurements, since the Q factor is determined locally in defect-free regions. However, we emphasize that structural disorder may play a critical role in extended polariton systems, where it can strongly affect coherence and transport. Global structural disorder can be reduced by optimizing growth conditions and strain management. For example, phosphorus incorporation has been reported to suppress crosshatch patterns associated with misfit dislocations \cite{Cilibrizzi2014, Zajac2012_3}. Additional approaches include optimized buffer layers, growth temperature tuning, reduction of thermal gradients, and the use of dislocation filtering layers.

Finally, the most straightforward way to increase the intrinsic cavity Q factor ($Q_{\mathrm{theor}}$) is to increase the number of DBR pairs, as illustrated in \Cref{fig:Qfac}. As shown above, for the present level of losses, the Q factor is expected to approach a limiting value of $Q \sim 5 \times 10^5$ with increasing number of DBR pairs, evaluated under the same detuning and spectral position of the photon mode as considered above.  However, this approach is ultimately limited by the accumulation of non-radiative losses, leading to a saturation of the Q factor at large mirror counts.

\begin{figure}[t]
  \centering
  \includegraphics[width=\linewidth]{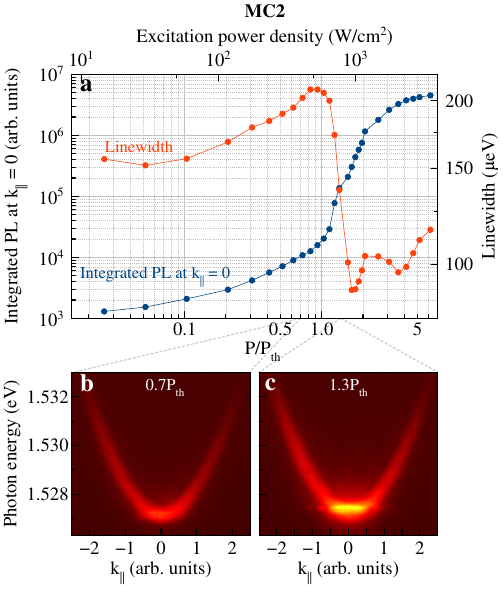}
  \caption{(a) Integrated PL intensity at $k_{\parallel}=0$ (blue) and linewidth of the polariton mode (red) as functions of the CW pump power density. (b) Angle-resolved PL spectrum below the threshold ($0.7 P_{\mathrm{th}}$) and (c) above the threshold pump density ($1.3 P_{\mathrm{th}}$).}
  \label{fig:MC35_PL}
\end{figure}

\textbf{Polariton lasing.}  
The threshold behavior of the polariton laser in the MC2 sample was analyzed. \cref{fig:MC35_PL}(a) shows the dependence of the integrated PL intensity at  $k_{\parallel}=0$ and the linewidth of the polariton mode on the CW pump power density. The threshold power density was determined from the linewidth behavior as the intersection of two linear fits: one describing the gradual linewidth increase at low excitation powers, and the other capturing the rapid linewidth narrowing above threshold. At the threshold pump density of $P_{\mathrm{th}} \approx \SI{570 \pm 90}{\watt\per\centi\meter\squared}$, a sharp increase in the PL intensity is observed, accompanied by a pronounced narrowing of the polariton linewidth, indicating the onset of polariton lasing. This value was obtained under continuous-wave excitation at the energy corresponding to the first minimum of the DBR stop band ($E_{\mathrm{pump}} \approx \SI{1.625}{\electronvolt}$), at a detuning of $\delta \approx -9~\mathrm{meV}$ and a temperature of 3.5~K. The uncertainty in the threshold determination is primarily governed by the finite width of the transition region and the determination of the excitation spot size. It is important to emphasize that the threshold is reached while the system remains in the strong-coupling regime, as confirmed by the angle-resolved dispersion measured at the threshold pump density. Below the threshold [\Cref{fig:MC35_PL}(b)], the characteristic dispersion of the lower polariton branch is observed. Above the threshold [\Cref{fig:MC35_PL}(c)], a strong enhancement of the emission near $k_{\parallel}=0$ appears, together with a small blue shift of the emission peak, which is a well-known signature of polariton lasing. The obtained threshold value is low for a GaAs/AlGaAs system at cryogenic temperatures under continuous-wave pumping.

 \section{Comparison with state-of-the-art microcavities}
 \label{sec:comparison}
 
In this section, the parameters of the MC2 microcavity are compared with those of representative planar microcavities reported in the literature, including both GaAs-based structures and microcavities based on other material platforms. The reported values of the quality factor $Q$ and the polariton lasing threshold $P_\mathrm{th}$ are taken from the original publications. It should be emphasized that a direct quantitative comparison of threshold values is complicated by the strong dependence of $P_\mathrm{th}$ on a number of experimental parameters, including the excitation spot size and shape \cite{Askitopoulos2013}, exciton-photon detuning \cite{Wertz2009, Schmutzler2013}, the presence and type of trapping potential, the excitation regime (CW, quasi-CW, or pulsed) \cite{Steger2017}, excitation photon energy, temperature \cite{Tsotsis2012, Tempel2012}, as well as the specific definition and extraction procedure used for the threshold. Similar considerations apply to the reported $Q$ factors, which may be determined from the cavity photon linewidth, from the polariton linewidth at a given detuning, or indirectly via lifetime measurements.

\subsection{GaAs-based microcavities}

 \begin{table}[b]
\caption{Comparison of representative planar GaAs-based microcavities reported by leading growth groups. The values of the quality factor $Q$ and the polariton lasing threshold $P_\mathrm{th}$ are taken from the original publications and should be regarded as approximate. Differences in measurement conditions and definitions of $Q$ and $P_\mathrm{th}$ should be taken into account.}
\label{tab:GaAs_comparison}
\begin{ruledtabular}
\begin{tabular}{lcc}
Reference & $Q$ & $P_\mathrm{th}$ (kW/cm$^{2}$) \\
\hline
Tsotsis \textit{et al.} \cite{Tsotsis2012} & $1.6\times10^{4}$ & 0.64 \\
Wertz \textit{et al.} \cite{Wertz2009} & $1.2\times10^{4}$ & 3 \\
Tinkler \textit{et al.} \cite{Tinkler2015} & $1.5\times10^{4}$ & 1.5 \\
Tempel \textit{et al.} \cite{Tempel2012} & $5\times10^{3}$ & 1.2 \\
Steger \textit{et al.} \cite{Steger2017} & $3.2\times10^{5}$ & 0.009--200 \\
\textbf{This article}            & $\boldsymbol{5.4\times10^{4}}$ & \textbf{0.57} \\
\end{tabular}
\end{ruledtabular}
\end{table}

The MC2 microcavity is compared with representative planar GaAs-based microcavities reported by leading growth groups (Table~\ref{tab:GaAs_comparison}). Within this context, planar GaAs-based microcavities reported in the literature typically exhibit quality factors in the range $Q \sim 5\times10^{3}$--$2\times10^{4}$, with polariton lasing thresholds on the order of $0.5$--$5~\mathrm{kW/cm^{2}}$ under nonresonant excitation with a focused spot.

More recently, the growth of GaAs-based microcavities with substantially higher quality factors, reaching up to $Q \sim 3\times10^{5}$, has been demonstrated. It is important to note, however, that even within this series of ultrahigh-$Q$ samples the reported threshold power densities span several orders of magnitude. In particular, threshold values ranging from $0.009$ to $200~\mathrm{kW/cm^{2}}$ have been reported for the same microcavity under different excitation conditions, emphasizing the strong dependence of the threshold on excitation geometry, detuning, and pumping regime rather than on the quality factor alone.

Against this background, the MC2 microcavity demonstrates a quality factor ($Q \approx (5.4 \pm 0.1)\times10^{4}$) that is comparable to that of state-of-the-art planar GaAs-based structures grown using conventional AlGaAs alloys, while simultaneously exhibiting a low polariton lasing threshold ($P_\mathrm{th} \approx (0.57 \pm 0.09)~\mathrm{kW/cm^{2}}$) under spot excitation. The comparison presented here demonstrates that the digital-alloy approach based on short-period GaAs/AlAs superlattices enables competitive performance relative to conventional ternary-alloy microcavity designs.

\subsection{Other material system}

Table~\ref{tab:material_comparison} summarizes representative results on polariton condensation demonstrated in microcavities based on a range of alternative material platforms, including CdTe, GaN, ZnO, organic semiconductors, halide perovskites, and transition-metal dichalcogenides (TMDCs). Despite significant differences in material properties, such as exciton binding energy and oscillator strength, polariton condensation has been successfully achieved in all these systems. A key advantage of several wide-bandgap materials and strongly bound excitonic systems is the possibility of realizing polariton lasing at room temperature, with threshold power densities on the order of a few $\mathrm{W/cm^2}$. These features make such platforms particularly attractive for integrated polaritonic devices operating under ambient conditions. However, these platforms often suffer from drawbacks such as low $Q$ factors, limited device stability, and challenges in fabrication control.

In contrast, GaAs-based microcavities typically require cryogenic temperatures due to the relatively low exciton binding energy in GaAs. Nevertheless, they offer a set of unique advantages that are crucial for fundamental studies of polariton physics, including ultra-high optical quality (high $Q$ factors), low inhomogeneous broadening, and precise control over exciton--photon detuning and cavity design. As a result, GaAs-based platforms remain the system of choice for investigating coherent and non-equilibrium phenomena in polariton condensates, such as superfluidity and quantized vortices.

\begin{table}[t]
\centering
\caption{Comparison of microcavities based on different material platforms. The values of the quality factor $Q$ and the polariton lasing threshold $P_\mathrm{th}$ are taken from the original publications and should be regarded as approximate. Differences in measurement conditions and definitions of $Q$ and $P_\mathrm{th}$ should be taken into account.}
\label{tab:material_comparison}
\begin{ruledtabular}
\begin{tabular}{l l c c}
Material & Reference & $Q$ & $P_\mathrm{th}$ (kW/cm$^{2}$)  \\
\hline
CdTe         & Kasprzak \textit{et al.} \cite{Kasprzak2006}  & $2.1\times10^{3}$   & 0.68 \\
GaN          & Jayaprakash \textit{et al.} \cite{Jayaprakash2017} & $1.8\times10^{3}$   & 0.0045 \\
ZnO          & Lu \textit{et al.} \cite{Lu2012}               & $2.2\times10^{2}$    & 32 \\
Organic      & Dietrich \textit{et al.} \cite{Dietrich2016}     & $7\times10^{2}$    & 300 \\
Perovskite   & Zou \textit{et al.} \cite{Zou2025}               & $10^{3}$      & 0.0004 \\
TMDC         & Fan \textit{et al.} \cite{Fan2024}               & $4.8\times10^{2}$    & 0.006 \\
\textbf{GaAs}         & \textbf{This article}            & $\boldsymbol{5.4\times10^{4}}$ & \textbf{0.57} \\
\end{tabular}
\end{ruledtabular}
\end{table}

\section{Conclusion}

We have demonstrated and investigated an approach for the growth of high-$Q$ planar GaAs/AlGaAs microcavities in which the high-index DBR layers are implemented as digital alloys based on short-period GaAs/AlAs superlattices. This design leads to an order-of-magnitude reduction of interface roughness, as confirmed by AFM measurements, enables precise control of both the λ/4 optical thickness and the effective Al content, suppresses the propagation of threading dislocations, and allows efficient tuning of background absorption. We demonstrated that selecting the appropriate superlattice period and duty cycle is essential for minimizing impurity-related absorption, which would otherwise limit the microcavity $Q$ factor. In addition, the use of digital alloy significantly simplifies the technological growth process compared to ternary alloys. Altogether, these advantages lead to an increased microcavity $Q$ factor, a key parameter governing the polariton lifetime and lasing threshold.

In the optimized MC2 microcavity, we obtained a low polariton-lasing threshold of $P_{\mathrm{th}} \approx (570 \pm 90)~\mathrm{W/cm^2}$ and a high $Q$ factor $Q_{\mathrm{expt}} \approx (5.4 \pm 0.1) \times 10^4$. This experimental value is nearly twice the theoretical estimate derived within the model of an equivalent ternary alloy. This indicates that accurate modeling of SPSL-based microcavities requires using the correct refractive index of the thin GaAs layers, which must account for the modified electronic density of states in the SPSL, including quantum-confinement and excitonic effects.

A semi-quantitative analysis of the loss mechanisms shows that, in the optimized structures, the dominant contributions arise from disorder-induced scattering and residual absorption, which are both reduced to the level of a few $\mu\mathrm{eV}$. Using these estimates, we obtain a good agreement between the calculated linewidth and the experimentally measured value, indicating that the present structures operate close to the disorder-limited regime.

Finally, we show that, at the present level of losses, further increase of the Q factor is expected to be limited, leading to saturation at $Q \sim 5 \times 10^5$ with increasing number of DBR pairs. While absorption losses can be further reduced by improving material purity and growth conditions, further progress will require control over the lateral correlation properties of structural disorder.

\begin{acknowledgments}
The growth and modeling of the structures were supported by the Russian Science Foundation (Grant No.~19-72-20039). The experimental studies were supported by St. Petersburg State University (Grant No.~125022803069-4). The MBE growth and AFM characterization were performed using the facilities of the Research Park of St. Petersburg State University “Center for Nanofabrication of Photoactive Materials (Nanophotonics)”. The Nomarski microscopy studies were carried out using the equipment of the Research Park of St. Petersburg State University “Interdisciplinary Center for Nanotechnology”.
\end{acknowledgments}

\appendix

\section{Molecular Beam Epitaxy}

The series of samples investigated in this work was grown by molecular beam epitaxy using an SVTA MBE 35–3 system (SVT Associates, Inc.), which provides ultrahigh-vacuum conditions with a base pressure of approximately $8\times10^{-11}~\text{Torr}$. GaAs(100) wafers with a small miscut ($\SI{0.06}{\degree}$) were used as substrates, ensuring two-dimensional growth. Before the microcavity growth, each substrate underwent a preheating procedure (5~hours at \SI{400}{\celsius}) to remove adsorbed water and hydrocarbons, followed by oxide desorption (15~minutes at \SI{600}{\celsius} under an $\mathrm{As}_4$ flux). After this treatment, a GaAs buffer layer of approximately \SI{1000}{\nano\meter} was deposited to form an atomically smooth surface suitable for subsequent epitaxy. In the middle of this buffer, a technological short-period GaAs/AlAs superlattice was grown to suppress the propagation of threading dislocations originating from the substrate.

\begin{figure*}[t]
  \centering
  \includegraphics[width=\linewidth]{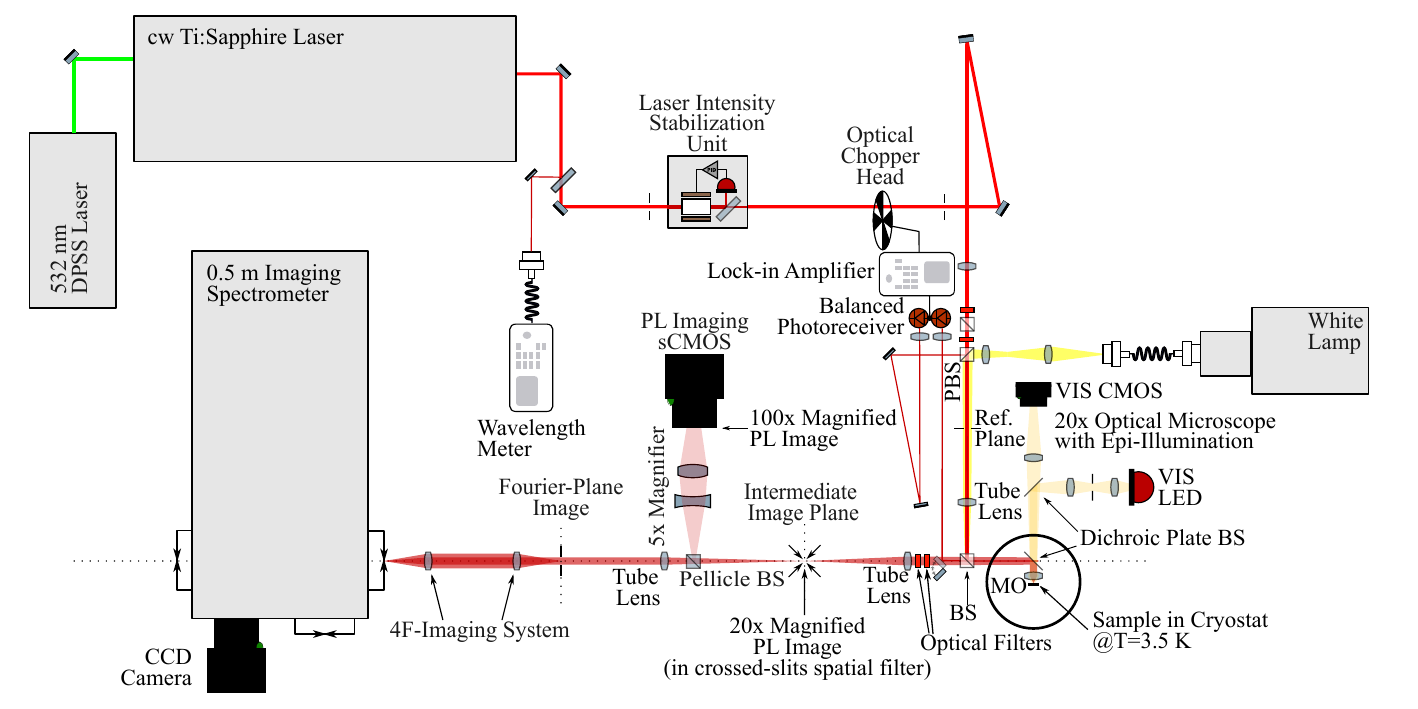}
  \caption{Experimental setup.}
  \label{fig:setup}
\end{figure*}

The growth was carried out at a substrate temperature of about \SI{580}{\celsius}, which represents a compromise for GaAs/AlAs epitaxy. The V/III beam equivalent pressure (BEP) ratio was maintained at about 2:1. During DBR growth, the average growth rate, determined from RHEED oscillations, was roughly one monolayer per second ($\sim$\,\SI{0.2827}{\nano\meter\per\second}), whereas for the cavity region the growth rate was reduced to 0.5~monolayers per second. The BEP values were also decreased by a factor of two for the cavity region while keeping the same 2:1 ratio. Two stable Al and Ga effusion cells were used during DBR growth, while the cavity layers were grown using the purest available Al source and a separate high-purity Ga source. The growth of a high-$Q$ microcavity may exceed 10~hours, requiring growth-rate stability better than 2\%. To monitor the growth rate, an \textit{Accuflux} module (SVT Associates, Inc.) based on absorption spectroscopy of molecular beams was employed. This allowed real-time, continuous monitoring of the metal fluxes and their correction by adjusting the temperatures of the effusion cells. A further advantage of using \textit{Accuflux} is that growth could be performed over several consecutive days: the module’s readings enabled accurate restoration of flux values at the beginning of each new growth session without the need for RHEED-based recalibration.

To obtain heterointerfaces with minimal roughness, short growth interruptions of about 10~seconds were introduced after each GaAs layer, while a longer 30-second interruption was applied after each quantum well. During these interruptions, the Ga flux was shuttered, and in the presence of an arsenic overpressure, surface Ga atoms, owing to their high mobility, relaxed and rearranged into a more ordered configuration. During DBR growth, the substrate was rotated at approximately $\SI{20}{rpm}$ to improve the lateral uniformity of the layer thicknesses across the wafer. As a result, only a small radial gradient in the DBR thickness remained in the structure. In contrast, during growth of the cavity, the substrate rotation was stopped, resulting in a detuning gradient of approximately $\sim$\,\SI{2}{\milli\electronvolt\per\milli\meter} across the sample, corresponding to a total detuning variation of about $\sim$\,\SI{92}{\milli\electronvolt} over the full \SI{46}{\milli\meter} wafer.

\section{Optical modeling of the structures.} 

To simulate the reflectance spectra and the electromagnetic field distribution in the microcavities, we employed a transfer-matrix method. The heterostructure was represented as a sequence of planar layers characterized by a complex refractive index 
\[
\tilde{n}_i(\lambda, T) = n_i + i k_i 
\]
and a thickness $d_i$. For every layer and every interface between adjacent layers, individual transfer matrices $M_i^{\mathrm{layer}}$ and $M_i^{\mathrm{interface}}$ were constructed, both depending on the refractive indices and thicknesses of the corresponding layers.

We explicitly include the excitonic contribution to the complex dielectric response in layers containing quantum wells. This contribution is described by a dedicated matrix $M^{\mathrm{QW}}$, fully consistent with the nonlocal dielectric-response formalism \cite{Ivchenko2005, Kavokin2017}. The matrix $M^{\mathrm{QW}}$ incorporates the intrinsic excitonic parameters: the radiative ($\Gamma_0$) and nonradiative ($\Gamma$) linewidth broadenings, the exciton transition frequency ($\omega_0$), and the phase of the excitonic contribution ($\phi$). These parameters were determined from experiments performed on single GaAs quantum wells of different thicknesses grown by molecular beam epitaxy in our laboratory \cite{Bataev2022}. By sequentially multiplying the matrices corresponding to all layers and interfaces, we obtained the total transfer matrix $M^{\mathrm{result}}$, from which the amplitude reflection coefficient of the heterostructure was calculated.

A crucial part of the modeling is the accurate choice of the complex dielectric permittivity, which must correctly reproduce the dispersion and absorption in \AlGaAs{} layers for arbitrary aluminum composition, for temperatures between 1 and 300~K, and in the spectral range 700--900~\si{\nano\meter}. In the absorption region of \AlGaAs{}, we employed the parametrization from Ref.~\cite{Aspnes1986}, while in the transparency region we used the polynomial model from Ref.~\cite{Gehrsitz2000}. Both parametrizations correspond to room-temperature data. For lower temperatures, the refractive-index dispersion was adjusted by shifting the spectral dependence in accordance with the temperature-induced shift of the band-gap energy.

\section{Optical characterization of the samples and measurement of the microcavity $Q$ factor}

Optical characterization of the samples was performed using reflectance and PL spectroscopy (\Cref{fig:setup}). The samples were mounted in a continuous-flow helium cryostat providing a base temperature of \SI{3.5}{\kelvin}. A broadband white-light source was used to record reflectance spectra, while PL was excited by a tunable cw Ti:Sa laser operating in the wavelength range 700–850~\si{\nano\meter}. The laser wavelength was monitored with a high-precision wavelength meter, and the optical power was controlled using a liquid-crystal-based laser intensity stabilizer followed by an optical attenuator and a polarizing beamsplitter. The excitation beam was focused onto the sample with a 20$\times$ microscope objective. For sample surface visualization, an amber-light episcopic illumination system and a CMOS camera were used.

The PL signal (or the reflected light in reflectance measurements) passed through a set of filters to suppress the pump beam and was focused by a tube lens onto the intermediate image plane (IP). A tunable crossed-slit pinhole was placed at the IP and served as a spatial filter. After the slit, the light was split into two paths. The first beam was directed onto an sCMOS camera to visualize a magnified image of the IP. The angle-resolved PL image was formed in the Fourier image plane by a second tube lens. It was then refocused onto the entrance slit of a 0.5-m imaging spectrometer using a 4$F$ imaging system. This configuration made it possible to record spectra in both real space and reciprocal space by translating one of the lenses out of the optical path. To record broadband spectra, we used a spectrometer with a spectral resolution of approximately \SI{35}{\micro\eV}, which allowed us to observe the DBR stop band in reflectance spectra and the polariton luminescence in PL. However, this resolution was insufficient for detecting the photon mode and polariton branches in the reflectance spectra of high-$Q$ microcavities, where the resonance linewidth can be much narrower than the spectrometer resolution. In particular, determining the microcavity $Q$ factor using \Eqref{eq:Q-def} was not feasible with spectra acquired using a standard spectrometer.

To overcome this limitation, we employed a tunable Ti:Sa narrow-linewidth ring laser that served as a high-precision spectroscopic probe. The laser frequency could be continuously tuned across the photon-mode spectral range ($\sim$800–820~\si{\nano\meter}) with a step of about $5\times10^{-5}$~\si{\nano\meter}, enabling detailed measurements of narrow resonances in high-$Q$ microcavities. To improve measurement stability, we used a balanced photodetection scheme. In this approach, the reflected signal from the sample was directed to one photodiode of the balanced detector, while a fraction of the incident laser light was used as a reference beam and detected by the second photodiode. The laser intensity was modulated by an optical chopper at \SI{1000}{\hertz}. The output of the balanced detector was fed into a lock-in amplifier synchronized with the chopper frequency. Subtracting the two photodiode signals suppressed laser power fluctuations. As a result, the balanced detection approach provided high sensitivity and enabled detection of resonance lines with FWHM on the order of \SI{1}{\micro\eV}.

\section{Disorder-induced losses in microcavities}

Thickness fluctuations and interface roughness of the constituent layers are closely related manifestations of structural disorder. After spatial averaging, both can be treated within the same disorder framework, namely as fluctuations of the local optical thickness. However, fluctuations in the DBR mirrors and in the cavity spacer affect the Q factor in qualitatively different ways.

Fluctuations in the DBR mirrors lead to phase decoherence of the reflected light. For small thickness fluctuations compared to the wavelength, the reduction in reflectivity can be estimated as
\begin{equation}
R = R_0 \exp\left[-\left(\frac{4\pi n \sigma}{\lambda}\right)^2 N\right],
\end{equation}
where $R_0$ is the reflectivity of the ideal DBR, $\sigma$ is the RMS roughness, $N$ is the number of layers, and $\lambda$ is the wavelength. In our structures, this contribution is very small, resulting in a broadening on the order of $\sim 0.2~\mu\mathrm{eV}$.

Fluctuations of the effective cavity spacer are more important. Here, by the effective cavity spacer we refer to the central cavity layer together with several adjacent DBR pairs that significantly contribute to the optical field distribution. These fluctuations directly affect the resonance energy of the photon mode:
\begin{equation}
\frac{\delta E_c(r)}{E_c} \approx \frac{\delta L_c(r)}{L_c}.
\end{equation}

The effect of cavity-thickness disorder depends critically on its lateral correlation length $\xi$ \cite{Duparre2002}. For very short-range disorder ($\xi \ll \lambda$), fluctuations are averaged out and the resulting losses are weak, while for disorder smooth on a scale much larger than the wavelength ($\xi \gg \lambda$), one expects predominantly inhomogeneous broadening.

Different lateral regions of the cavity resonate at slightly different photon energies, and the measured spectrum is broadened by the corresponding distribution of local resonances. For a \tlp\ cavity at $\lambda = 800$~nm and $n \approx 3.5$, the physical cavity thickness is $L_c \approx 350$~nm. This leads to an rms fluctuation of the local photon energy given by
\begin{equation}
\delta E_{\mathrm{RMS}} \approx E_c \frac{\sigma}{L_c}.
\end{equation}

Using $E_c = 1.55$~eV, one obtains $\delta E_{\mathrm{RMS}} \approx 0.4~(4)$~meV for $\sigma = 0.1~(1)$~nm. For a Gaussian distribution, the corresponding inhomogeneous linewidth (FWHM) is
\begin{equation}
\Delta E = 2\sqrt{2\ln 2}\,\delta E_{\mathrm{RMS}},
\end{equation}
i.e., approximately $0.9~(9)$~meV.

Thus, in the case of long-range disorder, one expects the quality factor to be limited to values of only a few thousand even for RMS roughness as low as $0.1$~nm.

\begin{figure}[b]
  \centering
  \includegraphics[width=\linewidth]{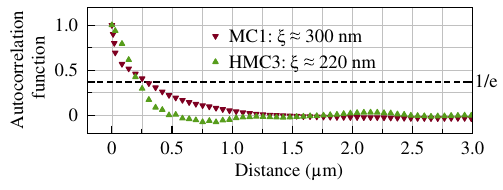}
  \caption{Autocorrelation function of the surface roughness extracted from AFM measurements for MC1 and HMC3 structures.}
  \label{fig:ACF}
\end{figure}

In our samples, AFM measurements (\Cref{fig:AFM}) indicate that the lateral scale of the roughness does not fall into either of the asymptotic limits. The measured autocorrelation function is used to estimate the lateral correlation length, yielding $\xi \approx 300$~nm for the MC1 microcavity, defined as the distance at which the autocorrelation function decays to $1/e$ of its maximum value (\Cref{fig:ACF}). This lateral scale is comparable to the optical wavelength in the material and therefore cannot be treated as either long-range disorder (leading to purely inhomogeneous broadening) or short-range disorder (leading to complete spatial averaging).

In this intermediate regime, the position of the photon mode is determined by the average cavity thickness, while the thickness fluctuations predominantly lead to scattering losses from the ideal cavity mode. In other words, the disorder does not simply induce a local energy shift, but also couples the fundamental cavity mode to finite-$k$ photonic states, thereby reducing the quality factor.

Within our analysis, this effect can be described as
\begin{equation}
\hslash \Gamma_{\mathrm{scat}} \sim E_c 
\int_{0}^{k_{\max}} 2\pi k
\left( \frac{2\pi n \sigma \xi}{\lambda} \right)^2 
\exp\left(-\frac{k^2 \xi^2}{4}\right) \, dk,
\end{equation}
which explicitly demonstrates the role of both the RMS roughness $\sigma$ and the lateral correlation length $\xi$. The integration must take into account the dispersion of the cavity photon mode and is therefore restricted to small $k$ values near the bottom of the parabolic dispersion.

As an order-of-magnitude upper-bound estimate for disorder-induced scattering losses, one may write
\begin{equation}
\hslash \Gamma_{\mathrm{scat}} \sim E_c \left( \frac{2\pi n \sigma}{\lambda} \right)^2.
\end{equation}
This estimate is consistent with the classical result of Bennett and Porteus~\cite{Bennett1961}.

Using this expression, we obtain the following upper-bound estimates: for $\sigma = 1$~nm, the corresponding broadening due to roughness-induced scattering is $\hslash \Gamma_{\mathrm{scat}} \approx 200~\mu\mathrm{eV}$, while for $\sigma = 0.1$~nm it is $\hslash \Gamma_{\mathrm{scat}} \approx 2~\mu\mathrm{eV}$.

\bibliography{HighQ_eng}

\end{document}